\documentclass[lettersize,journal]{IEEEtran}
\usepackage[switch]{lineno}

\usepackage{amsmath,amssymb,amsfonts}
\usepackage{algorithmic}
\usepackage{graphicx}
\usepackage{textcomp}
\usepackage{orcidlink}
\usepackage[table]{xcolor}
\usepackage{cite}

\usepackage{algorithm}
\usepackage{array}
\usepackage[caption=false,font=normalsize,labelfont=sf,textfont=sf]{subfig}
\usepackage{stfloats}
\usepackage{url}
\usepackage{verbatim}
\usepackage{booktabs}
\usepackage{hyperref}       % hyperlinks
\usepackage{url}            % simple URL typesetting
\usepackage{multirow}
\usepackage{pifont}
\usepackage{wrapfig}
\usepackage[most]{tcolorbox}
\usepackage{balance}

\begin{document}
\bstctlcite{IEEEexample:BSTcontrol}

\title{An Orchestrated Sandbox for Side-Channel Analysis of Cyber-Physical Systems Malware}
\author{Meet Udeshi\orcidlink{0000-0001-7297-0880}, \IEEEmembership{Graduate Student Member, IEEE}, 
Venkata Sai Charan Putrevu, 
Prashanth Krishnamurthy\orcidlink{0000-0001-8264-7972}, \IEEEmembership{Member, IEEE},
Ramesh Karri\orcidlink{0000-0001-7989-5617}, \IEEEmembership{Fellow, IEEE},
Farshad Khorrami\orcidlink{0000-0002-8418-004X}, \IEEEmembership{Fellow, IEEE}

\thanks{Received 16 August 2026; accepted 19 August 2026. Date of current version
19 October 2026. This work was supported in part by DOE National Energy
Technology Laboratory (NETL) under Grant DE-CR0000051 and Grant DECR0000017 and in part by NYSTAR under Grant C220160. This manuscript
was recommended for publication by J. Lee. (Corresponding author: Meet
Udeshi.)}
\thanks{Meet Udeshi, Prashanth Krishnamurthy, Ramesh Karri, and
Farshad Khorrami are with the Department of ECE, NYU Tandon School of
Engineering, Brooklyn, NY 11201 USA (e-mail: m.udeshi@nyu.edu).}
\thanks{Venkata Sai Charan Putrevu is with the Department of CSE, IIT Tirupati,
Tirupati 517619, India.}
\thanks{Digital Object Identifier 10.1109/LES.2026.3726537}
}

% The paper headers
\markboth{IEEE Embedded Systems Letters, VOL. 18, NO. 5, OCTOBER 2026}%
{Udeshi \MakeLowercase{\textit{et al.}}: An Orchestrated Sandbox for Side-Channel Analysis of Cyber-Physical Systems Malware}

\IEEEpubid{1943--0671~\copyright~2026 IEEE}

\maketitle

\begin{abstract}
Embedded devices powering cyber-physical systems (CPS) are increasingly internet-exposed, placing critical infrastructure under active threat. 
Dynamic analysis of embedded malware requires a sandbox that provides a realistic embedded environment,
captures system-level side channels for behavioral insight, and supports repeatable experiments. 
Existing Linux sandboxes fall short on all three: they target narrow tasks, run minimal environments, and only execute samples directly. 
Malware often withholds its true behavior, awaiting environmental triggers or command-and-control (C2) servers, or deploying anti-analysis techniques.
To address these issues, we present SaMOSA (sandbox for malware orchestration and side-channel analysis), with \emph{on-demand C2 response replay} via a \emph{replay formulation} tool to extract and serve C2 payloads, \emph{lifecycle-bound  orchestration} for environment customization, four synchronized  side-channels, and multi-architecture support. Among Linux sandboxes targeting CPS, only SaMOSA provides both on-demand C2 response replay and  lifecycle-bound orchestration, serving  payloads over  malware-initiated sessions for network-isolated execution without live infrastructure.
Through a Kinsing cryptominer case study on ARM64, we show that SaMOSA delivers the full C2 payload chain without live infrastructure, with synchronized channels resolving phase-wise mining behavior under live C2. Replay reproduces payload delivery but not stateful C2 sessions, a fidelity boundary that surfaced a hidden Tor fallback and 59 new IPs.
\end{abstract}


\section{Introduction}

Cyber-physical systems (CPS) increasingly rely on internet-exposed embedded devices running Linux, and the malware targeting those devices has grown accordingly: staging payloads, coordinating through command-and-control (C2) servers, and deploying anti-analysis techniques to evade scrutiny \cite{alhanahnah2018efficient, khorrami1}. 
For embedded malware, useful analysis has three requirements that are difficult to satisfy together. 
The sandbox must present a realistic environment, since samples that detect a minimal or virtualized environment withhold their true malicious behavior \cite{xie2021envfaker, krishnamurthy1};
it must capture synchronized, system-level side-channels for rich insights into execution behavior \cite{cozzi2018understanding, krishnamurthy2024tracking, krishnamurthy2, udeshi2};
and it must be repeatable, since malware analysis drives development of defenses \cite{alrawi2024sok}. 
When C2 infrastructure rotates domains or goes offline, samples that reached live infrastructure in one run may produce no behavior in the next, making repeatability difficult \cite{rossow2012prudent}. Some approaches target record-and-replay methods that replay entire execution traces or network packets \cite{dolangavittPanda2015}, however these fail to capture different behaviors of malware under different conditions.

Existing Linux sandboxes for CPS fall short of these requirements.
Most execute samples directly, without orchestration to customize the environment for triggering malicious behavior or to replay C2 responses on demand, so samples that depend on environmental triggers or live infrastructure never reveal their full behavior.
Many also run minimal buildroot-based images rather than full-featured operating systems which fail basic anti-analysis checks and further suppress malicious behavior.
We present SaMOSA, a sandbox designed around exactly these principles: orchestration for customization and triggering malware behavior, on-demand response replay for repeatable payload availability without live infrastructure, and synchronized multi-channel capture for comparative behavioral analysis.
Existing sandboxes share some of SaMOSA's components but none combine on-demand dynamic C2 response replay with lifecycle-bound orchestration, the  novel capabilities that  enable the environment-aware malware analysis workflows we demonstrate in this paper. Replay reconstructs payload content faithfully but not per-session C2 state, resulting in a divergence that surfaces new results (Section~\ref{sec:case_studies}).

The contributions of this paper are as follows:
1) \textbf{On-demand C2 response replay} along with a \textbf{replay formulation tool} that extracts C2 responses and serves them on-demand, preserving protocol correctness for payload delivery while freeing analysis from live infrastructure;
2) \textbf{Lifecycle-bound orchestration} for multi-architecture embedded Linux analysis, composing environment setup, host network mode, and external interactions (C2 servers, automated clients) into repeatable analyses;
3) \textbf{Two case-studies}, Kinsing cryptominer on ARM64 and CHAOS RAT on x86-64, demonstrating on-demand C2 replay for network-isolated multi-stage payload delivery, surfacing new behaviors not exposed by stub-based or live infrastructure runs \cite{udeshi1}.

\IEEEpubidadjcol
\noindent
\textbf{Threat Model:}
We consider a Linux ELF malware sample intended for embedded CPS devices that may coordinate behavior through remote C2 infrastructure.
The analyst's goal is to observe and reproduce the malware's full behavior in a
controlled environment, with or without live infrastructure, enabling deep iterative
analysis across different configurations and OS environments.
Analog sensor/actuator paths and physical-layer leakage are out of scope. 
SaMOSA targets system-level behavioral analysis of Linux malware on embedded CPS-class devices, where the emulated ARM64 target and host-side telemetry match the deployment and defense context of \cite{krishnamurthy2024tracking}. Field devices cannot be instrumented or infected in situ, so faithful emulation of the target architecture is the practical route to observing malware behavior. The same hooks that supply environmental triggers can also attach a plant simulation, positioning SaMOSA to extend beyond malware-triage to closed-loop CPS validation.

\section{Related Work} \label{sec:related}

Cuckoo~\cite{CuckooSandbox} and CAPE~\cite{CAPE} offer pre/post hooks but are
Windows-centric, with limited multi-architecture Linux support and no
lifecycle-bound host network control; DRAKVUF \cite{drakvuf} targets kernel-event
tracing on x86 Xen; Ansible and Terraform provision infrastructure but provide
neither network isolation nor side-channel capture, and are instead invokable
from SaMOSA hooks. Padawan~\cite{cozzi2018understanding} traces system calls in a
QEMU VM but is unavailable. LiSa~\cite{uhricek2020lisa} (SystemTap) and
ELFEN~\cite{ELFEN} capture events from inside small buildroot images rather than
a full-featured OS, with filters targeting specific events. None triggers a
sample's full behavior across architectures without on-demand replay or
orchestration.
FakeNet-NG and INetSim serve only stub responses. Conventional packet replay like
\texttt{tcpreplay}/\texttt{tcprewrite} and
\texttt{fragroute} manipulate recorded frames, while GoReplay replays captured
requests to a live server, whereas isolated analysis requires serving responses
to an uncooperative client initiating fresh sessions. SaMOSA serves real
extracted C2 payloads keyed by request URL, so a sample cycling through C2 IPs is
still served correctly. Unlike whole-system
record-and-replay~\cite{dolangavittPanda2015}, which deterministically re-traces a fixed
execution, SaMOSA drives a live execution, allowing new behaviors to surface~\cite{blackReanimatingHistoricMalware2021}.

\section{SaMOSA Sandbox}

SaMOSA is built on top of virtual machine (VM) emulation using QEMU, supporting ARM64 and x86-64.
We operate QEMU in full-system emulation mode and, unlike existing sandboxes, we build VM images from the full-featured Ubuntu 20.04 operating system (OS) instead of a minimal buildroot OS, providing a 
realistic environment that defeats minimal-rootfs and buildroot heuristics, though QEMU artifacts remain detectable and can be countered through the setup hook.
% For x86-64, we run the same machine configuration as the host CPU and use an NVMe disk device.
For x86-64, we use host CPU configuration and an NVMe disk device.
For ARM64, we run the \texttt{virt} machine with a Cortex A72 CPU to mimic Raspberry Pi-class hardware, representative of the gateways and field controllers that mediate CPS sensing and actuation, and attach a \texttt{virtio} block device.
SaMOSA can incorporate custom VM images and OSes.

\begin{figure}[tpb]
    \centering
    \includegraphics[width=\linewidth]{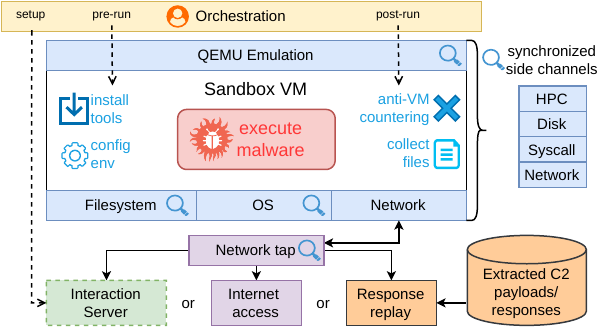}
    \caption{Execution pipeline of the SaMOSA sandbox, with automated orchestration for environment customization, on-demand response replay for repeatable analysis, and time-synchronized side-channel capture.}
    \label{fig:sandbox}
\end{figure}

SaMOSA captures four time-synchronized system-level side channels: hardware performance counters (HPC), disk activity, system calls, network activity.
We use \textit{side channel} to mean execution signals collected outside the malware's interference; system calls are the exception, requiring in-VM capture.
Side channels except system calls are captured on the host to avoid detection  or tampering by malware anti-VM checks. This mirrors CPS monitoring practice, where constrained field devices are observed externally and the resulting behavioral traces feed runtime anomaly detectors \cite{krishnamurthy2024tracking}.
System calls are captured using \texttt{sysdig}
running inside the sandbox.
HPC measurements are captured using \texttt{perf} on the  QEMU process, which is less accurate than guest-side measurement. However, since QEMU executes translated guest code, host-based HPC measurements for the QEMU process capture the guest computation trends.
Disk access measurements are generated via QEMU tracing of NVMe or virtio devices.
Network packets are captured using \texttt{tcpdump} via a network bridge and tap interface that isolates the sandbox network from the host. 
SaMOSA provides configurable connections and routing via IP tables so that the user can choose between full internet access, on-demand response replay, or a customized interaction server.
At VM startup, host and guest timestamps are recorded and used to normalize each
side-channel's timestamps to a common zero-based timeline, aligning all four channels
for direct comparison. 
Fig.~\ref{fig:sandbox} shows SaMOSA's execution pipeline.
\\

\noindent \textbf{Orchestration:}
SaMOSA's orchestration framework allows the user to customize the sandbox as desired for particular malware, such as to start custom interactions servers, install tools, setup environment configurations, counter anti-VM techniques, or copy malware-generated files, as shown with examples in Fig.~\ref{fig:sandbox}.
We define four hook points as follows: ``setup'' before the VM is setup and booted, ``pre-run'' before measurement starts and the binary is executed, ``post-run'' after binary is halted, and ``shutdown'' after the VM is powered off.
In this manner, the execution environment can be tailored to specific malware requirements, for example, by setting environment variables or modifying the firewall configuration in the sandbox before execution. Once an orchestration is set up, the  sample can be run repeatedly as the execution and side-channel collection is end-to-end automated.
\\

\noindent \textbf{Network Emulation and On-Demand C2 Response Replay:}
SaMOSA isolates sandbox network traffic via a dedicated bridge and tap interface on the host,
enabling two distinct network modes: full internet access for initial payload discovery, and
FakeNet-emulated network isolation.
The key contribution is SaMOSA's \emph{response replay} capability, which bridges these two
modes. Conventional packet replay tools cannot address this (see Section~\ref{sec:related}).
SaMOSA instead replays payloads over \emph{fresh} sessions initiated by the malware itself, preserving
TCP protocol correctness and eliciting genuine behavior.

\begin{figure}[htpb]
    \centering
    \includegraphics[width=\linewidth]{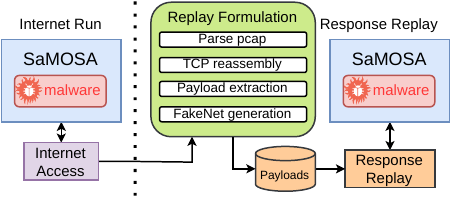}
    \caption{On-demand response replay pipeline with replay formulation.}
    \label{fig:response_replay}
\end{figure}

This is enabled by the \emph{replay formulation tool} (Fig.~\ref{fig:response_replay}), a four-stage
pipeline that: (i) parses raw \texttt{pcap} captures from the internet-enabled run,
(ii) performs TCP stream reassembly to recover application-layer exchanges,
(iii) extracts binary C2 response payloads per request URL and C2 address, and
(iv) generates a FakeNet configuration and FakeNet plugin that maps each URL to its corresponding payload.
In subsequent runs, FakeNet intercepts outbound requests from the malware and
the generated FakeNet plugin serves the matched payloads on demand, without any live infrastructure. 
The sandbox remains fully air-gapped, making the analysis network-isolated and repeatable.
This also enables \emph{retrospective analysis} with new defense techniques being tested on malware samples
whose C2 infrastructure is no longer available. The tool currently only supports HTTP. TLS protocols can be implemented with an MITM proxy, left for future work.
\\

\noindent \textbf{Performance Comparison:}
Table~\ref{tab:resource_consumption} benchmarks the overhead introduced by side-channel
monitoring, measured as the resource consumption difference between a full monitoring run
and a baseline run with monitoring disabled averaged across three runs, for both SaMOSA and LiSa. 
The benchmark workload is a C program that performs arithmetic computations and generates periodic network traffic.
We focus our comparison against LiSa since Padawan is not publicly available, and disabling ELFEN's monitoring would require significant re-engineering of its containerized execution pipeline.

\begin{table}[htpb]
\centering
\scriptsize
\caption{Overhead of side channel monitoring.}
\label{tab:resource_consumption}
\begin{tabular}{lccc}
    \toprule
    \; & \textbf{Arch}&  \textbf{CPU} & \textbf{Mem} \\
    & & (\%) & (MB) \\
    \cmidrule{2-4}
    \multirow{2}{0.07\linewidth}{SaMOSA} & x86-64 & -0.03	& +78.19 \\
                            & ARM64 & +0.24  & +31.95 \\
    \cmidrule{2-4}
    \multirow{2}{0.07\linewidth}{LiSa} & x86-64 & -0.04 & +41.33 \\
                                      & ARM64 & +1.22 & +51.87 \\
    \bottomrule
    \end{tabular}
\end{table}

Despite capturing four behavioral channels, running a full-featured OS, and providing
orchestration, SaMOSA's host CPU overhead is comparable to LiSa for x86-64 (negative values are within measurement noise) and only
$0.98\%$ higher for ARM64.
Host memory overhead is $+78$MB for SaMOSA versus $+41$MB for LiSa on x86-64, and
lower for SaMOSA on ARM64 ($+32$MB versus $+52$MB for LiSa).
SaMOSA's richer features come at negligible cost.

\begin{figure}[htpb]
    \centering
    \includegraphics[trim=30 20 30 25,width=\linewidth]{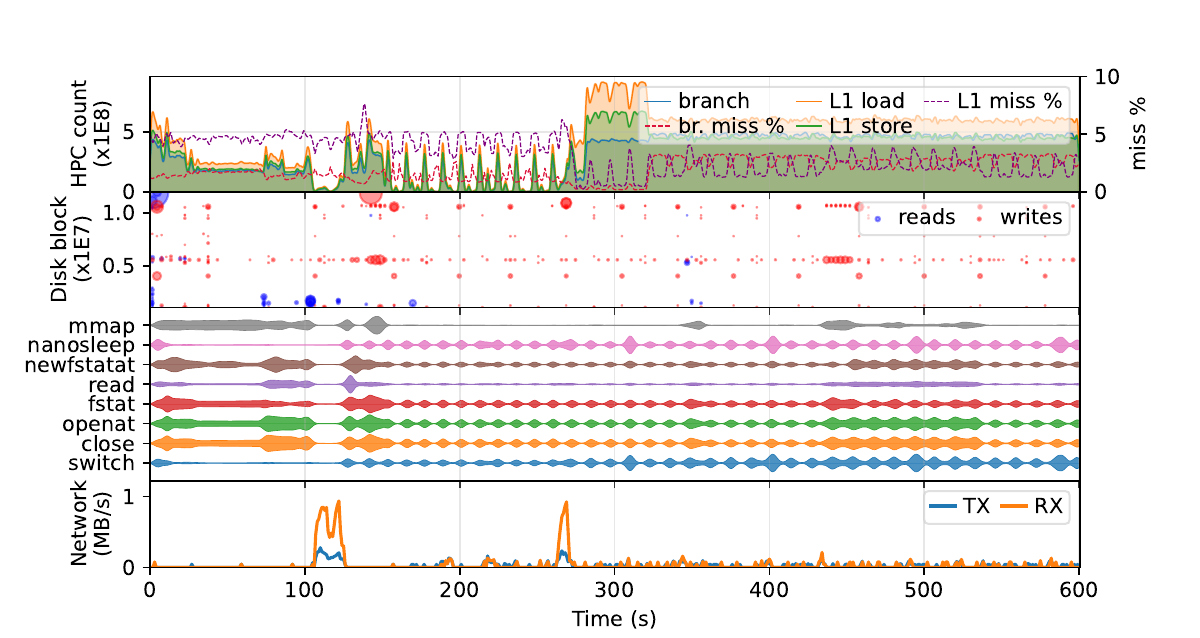}
    \caption{Execution of Kinsing on ARM64 when running with internet access.}
    \label{fig:kinsing_plot}
    % \vspace{-0.25em}
\end{figure}

\begin{figure}[htpb]
    \centering
    \includegraphics[trim=30 20 30 25, width=\linewidth]{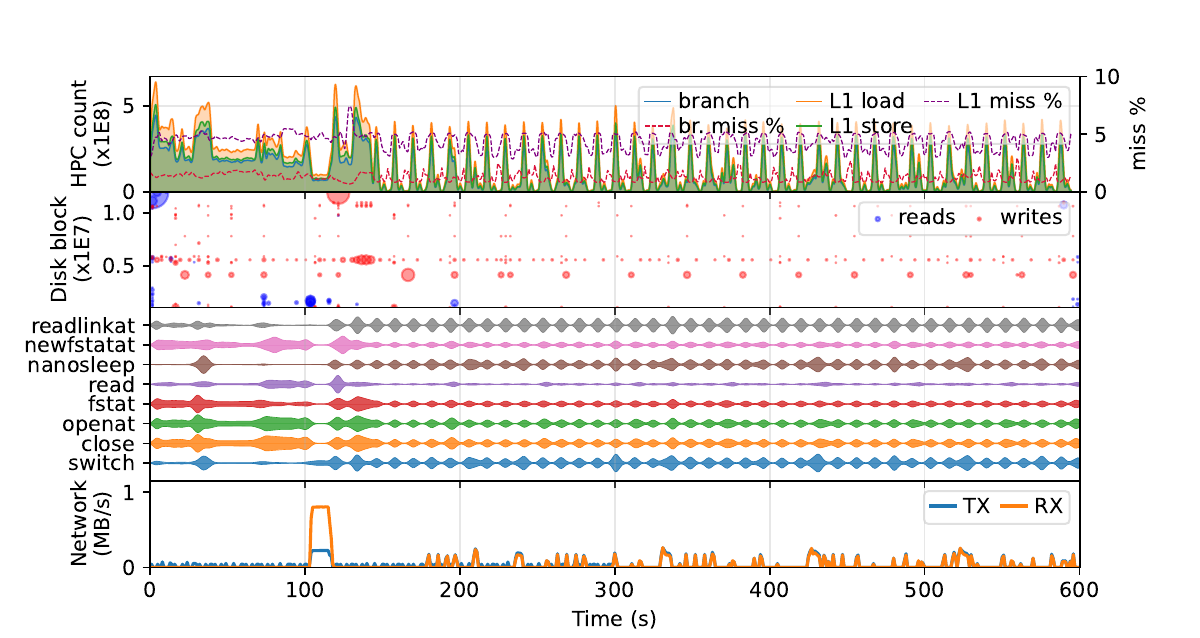}
    \caption{Execution of Kinsing when running with \emph{on-demand response replay}.}
    \label{fig:kinsing_plot_payload}
    % \vspace{-1.25em}
\end{figure}

\section{Case Study: Kinsing Cryptominer}
\label{sec:case_studies}

Kinsing is a cryptomining malware actively targeting CPS systems \cite{Kinsing2} via botnets, rootkits, and cryptocurrency miners. 
It demonstrates evasive anti-monitoring behavior by disabling system monitoring services such as SSH.
We execute the Kinsing sample on ARM64.
We add a ``post-run'' hook using the orchestration framework to counter the anti-monitoring behavior by re-enabling the SSH service.
\\

\noindent \textbf{1) Initial run:} Initially, we run the sample using the default FakeNet configuration that emulates network services with basic stub replies so that we capture accessed C2 IP addresses and URLs.
We observed HTTP requests for \texttt{kinsing\_aarch64}, \texttt{libsystem.so}, and \texttt{ce.sh} to C2 server IPs \texttt{78.153.XX.XX} and \texttt{80.64.XX.XX} (IP redacted for safety). %\footnote{These and other IPs are redacted for safety concerns.}.
The default FakeNet configuration only responds with basic stub replies so the sample terminates early.
With these insights, we trigger Kinsing again with full internet access to observe the download and execution of the real payloads.

\noindent\textbf{2) Internet access run:}
Fig.~\ref{fig:kinsing_plot} shows the execution activity when Kinsing is run with full internet access. For safety concerns, we provided internet access over a cellular network to avoid exposure to other devices on the university network, and only ran the sample inside the SaMOSA sandbox for 10 minutes to prevent long-term botnet or mining behavior \cite{rossow2012prudent}.
From 100 to 120 seconds, it downloads the ARM64-specific main binary (\texttt{kinsing\_aarch64}) and stealthy rootkit (\texttt{libsystem.so}).
The rootkit allows Kinsing to modify system behavior and hide its presence.
Between 260 to 270 seconds, it retrieves secondary payloads from additional IPs (\texttt{185.154.XX.XX}, \texttt{31.184.XX.XX}), demonstrating how malware behavior changes with full internet access.
At 270 seconds, Kinsing downloads the cryptominer payload \texttt{x\_arm} that is installed as a seemingly benign binary \texttt{/tmp/kdevtmpfsi} and executed in the background.
This initiates cryptocurrency mining, evident via the heavy HPC activity with notable spikes in memory stores and branch operations.
The bump in network traffic at 430 seconds along with subsequent disk write activity corresponds to the download of a persistence payload (\texttt{cron.sh}) from \texttt{78.153.XX.XX} which eliminates competing malware and establishes persistence. 
Kinsing also downloads an SSH scanner (\texttt{spre.sh}) here, which performs reconnaissance and demonstrates worm behavior. 
Another payload (\texttt{ce.sh}) was requested but never retrieved as its C2 IP was unreachable. 
After mining starts, periodic network activity reflects crypto seed exchanges and botnet heartbeat signals.
\\

\noindent \textbf{3) Response replay run:}
We parse the internet packet capture with the \emph{replay formulation tool} which generates a FakeNet configuration that describes redirections for each protocol accessed, specifically HTTP for Kinsing. The extracted payloads and corresponding URLs are packaged into a FakeNet plugin which serves the payloads on-demand when a HTTP request is redirected to it. We currently only match the HTTP URL but not the IP address, which helps when the sample cycles through multiple C2 IPs.
This yields the full set of stage-two binaries: the ARM64-specific \texttt{kinsing\_aarch64}, the rootkit \texttt{libsystem.so}, the cryptominer \texttt{x\_arm}, and the persistence and lateral-movement scripts \texttt{cron.sh} and \texttt{spre.sh}. 

The on-demand response replay allows for payload modification, a capability not provided by record-and-replay frameworks or other sandboxes.
We demonstrate this with \texttt{ce.sh}, which was never retrieved during the prior two runs
because its C2 IP was unreachable. Instead, we inject a script that records the process tree upon execution, revealing that \texttt{ce.sh} is called via a  \texttt{cron} entry.
Fig.~\ref{fig:kinsing_plot_payload} shows the side-channel activity for the response
replay run. Patterns align closely with the internet run until divergence at 270 seconds,
with subtle differences such as more uniform network speed due to local air-gapped
communication. The divergence is caused by a failure in C2 synchronization, so
cryptomining never starts despite \texttt{x\_arm} being delivered. This triggers a
resilience behavior not seen in prior runs: Kinsing cycles through multiple C2 IPs and
falls back to Tor, broadcasting a request to \texttt{rolibztiz....onion}. We observed
59 new IPs compared to the internet run, 9 for C2 and 50 for Tor.
Together, the three runs form an iterative analysis loop: stub-based C2 discovery, live payload retrieval, then network-isolated reproduction. Run 1 shows that without real payloads the sample terminates early, leaving no path to the cryptominer, rootkit, or lateral-movement stages.

% \vspace{-0.5em}

\section{Case Study: CHAOS Remote Access Trojan}

\begin{figure}[htpb]
    \centering
    \includegraphics[trim=15 20 15 0, width=\linewidth]{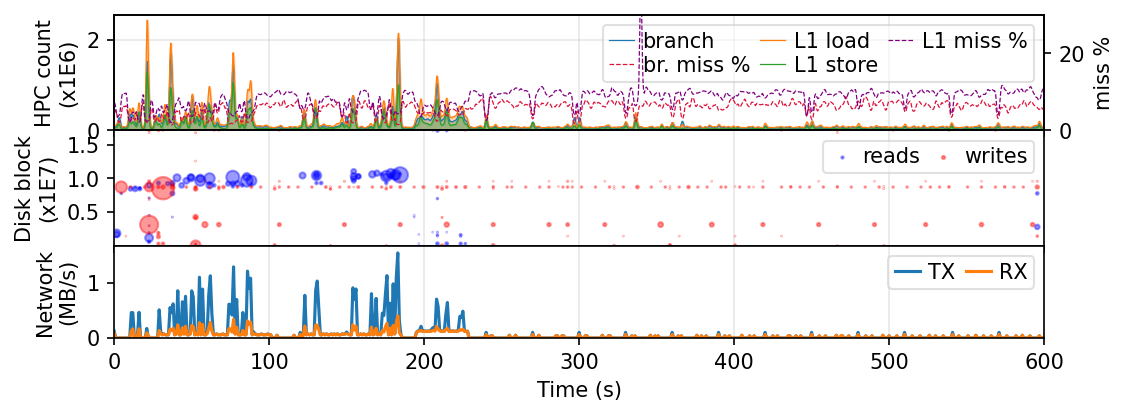}
    \caption{Execution of the CHAOS RAT on x86-64.}
    \label{fig:chaos_plot}
\end{figure}
CHAOS is a Golang-based RAT and
C2 framework misused for data exfiltration and botnets targeting CPS~\cite{chaos_acronis}.
Using SaMOSA's orchestration, we launched the C2 server 
and triggered an interaction script
that browsed folders, created a
privileged user, and downloaded system files.
Fig.~\ref{fig:chaos_plot} shows the execution activity. At 20 seconds, correlated spikes
in disk and network prompted deeper syscall analysis, revealing shell commands
that created a new user and modified permissions, each accessing low block addresses.
This demonstrates how cross-channel time-synchronized capture narrows the analysis to
 hotspots of malicious activity.

\section{Conclusion}

We presented SaMOSA, a Linux sandbox combining on-demand C2 response replay, lifecycle-bound orchestration, and synchronized four-channel capture. The Kinsing case study on ARM64 delivered the full stage-two payload chain in a network-isolated environment and surfaced a Tor fallback and 59 new IPs; the CHAOS RAT case study on x86-64 showed orchestration-driven C2 setup and automated client interaction for interactive malware. In both case studies, cross-channel synchronization narrowed the analysis to specific hotspots (mining  in Kinsing, credential access in CHAOS RAT) that a single channel would not have isolated. Two limitations remain: the replay formulation tool only supports HTTP; TLS requires an MITM proxy. The Kinsing replay also exposed a fidelity boundary: reproducing stateful C2 interactions requires protocol-aware replay, a natural extension of SaMOSA.
Beyond malware triage, SaMOSA is positioned to support closed-loop CPS validation against a simulated plant.

\IEEEtriggeratref{9}
\bibliographystyle{IEEEtran}
\bibliography{main}

@IEEEtranBSTCTL{IEEEexample:BSTcontrol,
CTLuse_forced_etal       = "yes",
CTLmax_names_forced_etal = "7",
CTLnames_show_etal       = "6" }

@misc{sysdig,
    title={Sysdig},
    author={Sysdig},
    howpublished={\url{https://github.com/draios/sysdig}}
}

@misc{CuckooSandbox,
    title={Cuckoo Sandbox},
    author={Cuckoo},
    howpublished={\url{https://github.com/cuckoosandbox/cuckoo}},
    note={Accessed: 2025-06-07},
}

@misc{CAPE,
    title={CAPE Sanbox},
    author={CAPE},
    howpublished={\url{https://github.com/kevoreilly/CAPEv2}},
    note={Accessed: 2025-06-07},
}

@inproceedings{cozzi2018understanding,
  title={Understanding linux malware},
  author={Cozzi, Emanuele and Graziano, Mariano and Fratantonio, Yanick and Balzarotti, Davide},
  booktitle={Symposium on Security and Privacy},
  pages={161--175},
  year={2018},
  organization={IEEE}
}

@misc{ELFEN,
    title={ELFEN Sandbox},
    author={ELFEN},
    howpublished={\url{ https://github.com/nikhilh-20/ELFEN}}
}

@misc{uhricek2020lisa,
  title={{LiSa}: multiplatform linux sandbox for analyzing {IoT} malware},
  author={Uhr{\i}cek, Daniel},
  year={2020},
  howpublished={\url{https://excel.fit.vutbr.cz/submissions/2019/058/58.pdf}}
}

@inproceedings{alhanahnah2018efficient,
  title={Efficient signature generation for classifying cross-architecture IoT malware},
  author={Alhanahnah, Mohannad and Lin, Qicheng and Yan, Qiben and Zhang, Ning and Chen, Zhenxiang},
  booktitle={2018 IEEE conference on communications and network security (CNS)},
  pages={1--9},
  year={2018},
  organization={IEEE}
}

@Article{krishnamurthy2024tracking,
AUTHOR = {Krishnamurthy, Prashanth and Rasteh, Ali and Karri, Ramesh and Khorrami, Farshad},
TITLE = {Tracking Real-Time Anomalies in Cyber-Physical Systems Through Dynamic Behavioral Analysis},
JOURNAL = {Journal of Cybersecurity and Privacy},
VOLUME = {6},
YEAR = {2026},
NUMBER = {2},
}

@misc{chaos_acronis,
author={Santiago Pontiroli and Gabor Molnar and Kirill Antonenko},
title={From open-source to open threat: Tracking Chaos RAT’s evolution},
year={2025},
url={https://www.acronis.com/en-us/cyber-protection-center/posts/from-open-source-to-open-threat-tracking-chaos-rats-evolution/},
note={Accessed: 2025-07-26}
}

@misc{Kinsing2,
    title={Kinsing Saltstack},
    author={Redcanary},
     year={2020},
    howpublished={\url{https://redcanary.com/blog/threat-intelligence/kinsing-malware-citrix-saltstack/}},
    note={Accessed: 2025-06-07},
}

@misc{alrawi2024sok,
      title={SoK: An Essential Guide For Using Malware Sandboxes In Security Applications: Challenges, Pitfalls, and Lessons Learned}, 
      author={Omar Alrawi and Miuyin Yong Wong and Athanasios Avgetidis and Kevin Valakuzhy and Boladji Vinny Adjibi and Konstantinos Karakatsanis and Mustaque Ahamad and Doug Blough and Fabian Monrose and Manos Antonakakis},
      year={2024},
      eprint={2403.16304},
      archivePrefix={arXiv},
      primaryClass={cs.CR},
      url={https://arxiv.org/abs/2403.16304}, 
}

@inproceedings{xie2021envfaker,
  title={Envfaker: A method to reinforce linux sandbox based on tracer, filter and emulator against environmental-sensitive malware},
  author={Xie, Chenglin and Guo, Yujie and Shi, Shaosen and Sheng, Yu and Chen, Xiarun and Li, Chengyang and Wen, Weiping},
  booktitle={International Conference on Trust, Security and Privacy in Computing and Communications},
  pages={667--677},
  year={2021},
  organization={IEEE}
}

@inproceedings{rossow2012prudent,
  title={Prudent practices for designing malware experiments: Status quo and outlook},
  author={Rossow, Christian and Dietrich, Christian J and Grier, Chris and Kreibich, Christian and Paxson, Vern and Pohlmann, Norbert and Bos, Herbert and Van Steen, Maarten},
  booktitle={Symposium on Security and Privacy},
  pages={65--79},
  year={2012},
  organization={IEEE}
}

@inproceedings{dolangavittPanda2015,
  title = {Repeatable {{Reverse Engineering}} with {{PANDA}}},
  booktitle = {Proceedings of the 5th {{Program Protection}} and {{Reverse Engineering Workshop}}},
  author = {{Dolan-Gavitt}, Brendan and Hodosh, Josh and Hulin, Patrick and Leek, Tim and Whelan, Ryan},
  year = 2015,
  pages = {1--11},
  publisher = {ACM},
}

@incollection{blackReanimatingHistoricMalware2021,
  title = {Reanimating {{Historic Malware Samples}}},
  booktitle = {Malware {{Analysis Using Artificial Intelligence}} and {{Deep Learning}}},
  author = {Black, Paul and Gondal, Iqbal and Vamplew, Peter and Lakhotia, Arun},
  year = 2021,
  pages = {345},
}

@misc{drakvuf,
    title={Black-box binary analysis system},
    author={DRAKVUF},
    howpublished={\url{https://drakvuf.com/}},
    note={Accessed: 2025-06-07},
}

@ARTICLE{khorrami1,
  author={Khorrami, Farshad and Krishnamurthy, Prashanth and Karri, Ramesh},
  journal={IEEE Design \& Test}, 
  title={Cybersecurity for Control Systems: A Process-Aware Perspective}, 
  year={2016},
  volume={33},
  number={5},
  pages={75-83},
  doi={10.1109/MDAT.2016.2594178}}

@ARTICLE{krishnamurthy1,
  author={Krishnamurthy, Prashanth and Khorrami, Farshad and Karri, Ramesh and Paul-Pena, David and Salehghaffari, Hossein},
  journal={IEEE Transactions on Information Forensics and Security}, 
  title={Process-Aware Covert Channels Using Physical Instrumentation in Cyber-Physical Systems}, 
  year={2018},
  volume={13},
  number={11},
  pages={2761-2771},
  doi={10.1109/TIFS.2018.2833063}}

@ARTICLE{krishnamurthy2,
author={Krishnamurthy, Prashanth and Karri, Ramesh and Khorrami, Farshad},
journal={ IEEE Micro },
title={{ Enabling Deep Visibility into VxWorks-Based Embedded Controllers in Cyber-Physical Systems for Anomaly Detection }},
year={5555},
volume={},
number={01},
ISSN={1937-4143},
pages={1-12},
doi={10.1109/MM.2025.3617807},
url = {https://doi.ieeecomputersociety.org/10.1109/MM.2025.3617807},
publisher={IEEE Computer Society},
address={Los Alamitos, CA, USA},
month=oct}

@INPROCEEDINGS{udeshi1,
  author={Udeshi, M. and Raz, M. and Krishnamurthy, P. and Karri, R. and Khorrami, F.},
  booktitle={2026 IEEE 32nd International Symposium on On-Line Testing and Robust System Design (IOLTS)}, 
  title={Sandbox-Enabled Digital Twin for Cyber-Physical Systems}, 
  year={2026},
  volume={},
  number={},
  pages={1-5},
  note={to appear.}}

@ARTICLE{udeshi2,
  author={Udeshi, Meet and Krishnamurthy, Prashanth and Karri, Ramesh and Khorrami, Farshad},
  journal={IEEE Transactions on Network and Service Management}, 
  title={Tamper-Proof Network Traffic Measurements on a NIC for Intrusion Detection}, 
  year={2025},
  volume={22},
  number={2},
  pages={2214-2224},
  doi={10.1109/TNSM.2024.3512180}}

\end{document}